\documentclass[preprint,12pt]{./elsarticle}

\usepackage{amssymb}

\usepackage{amsfonts,amsmath,amssymb,amsbsy,amsthm}

\usepackage{hyperref}
\usepackage{mathrsfs}
\usepackage{epstopdf}

\usepackage{graphicx}

\usepackage{subcaption}
\usepackage{multirow}

\usepackage{bbm} %indicator function

\begin{document}

\begin{frontmatter}

\title{A Critical Study of Baldelli and Bourdin's \emph{On the Asymptotic Derivation of Winkler-Type Energies From 3D Elasticity}}

\author{Kavinda Jayawardana\corref{mycorrespondingauthor}}
\cortext[mycorrespondingauthor]{Corresponding author}
\ead{zcahe58@ucl.ac.uk}

\begin{abstract}
In our analysis, we show that Baldelli and Bourdin's work \cite{Andres}  is only valid when describing the behaviour of a film bonded to an elastic pseudo-foundation, where Poisson's ratios of both bodies are in between $-1$ and $0$ or in between $0$ and $\frac12$ (where both Poisson's ratios are sufficiently away from $0$ and $\frac12$), and with an asymptotic condition that is different to what the authors present. We also show that, for all Poisson's ratios, the authors' phase diagram is four-dimensional and not two-dimensional. Also, due to the Poisson's ratio dependence, the asymptotic scalings that the authors present are insufficient to derive their proposed models. Furthermore, the authors' scaling of the displacement field implies that their method cannot be applicable to films (or strings) with planar loading, unless the normal displacement is zero. Finally, by deriving a Winkler foundation type solution for a plate supported by an elastic pseudo-foundation via the method implied by the authors, we show that the authors' method cannot be applied to plates due to the structure of the overlying body (i.e. the limits of integration of the plate) and the foundation (i.e. planar-stress free condition of the foundation), unless planar displacement field is identically zero. Despite the limitations of the authors' work, we highlight its strength by showing that unlike the classical derivation of the Winkler foundation equation, Baldelli and Bourdin's approach \cite{Andres} does not violate the volume conversation laws or violate the governing equations of mathematical elasticity.
\end{abstract}


\end{frontmatter}

%% main text
\section{Introduction}
\label{S:1}

Consider a situation where two elastic bodies that are bonded together, and for this case, one can easily model this problem with the simple use of the three-dimensional elastic equations. Now, consider the scenario where one of the elastic bodies is very thin in comparison to the other body, and planar in a Euclidean sense. Then the thin body can be approximated by a plate or a film, and such models are a frequently used in the of field of stretchable and flexible electronics. Applications of such models can be found in the field of conformal displays \cite{forrest2004path},  thin film solar cells \cite{choi2008polymers, crawford2005flexible,lewis2006material, pagliaro2008flexible}, electronic skins for robots and humans \cite{wagner2004electronic} and conformable electronic textiles \cite{ bonderover2004woven}. For such applications, the degree of the deformation of the electronic body can endure, before its basic functions (i.e. conductivity, transparency or light emission) are adversely affected, is immensely important. However, design and process engineers who are working on the implementation of flexible electronics often lack confidence due to a lack of understanding or a lack of input data for reliable modelling tools \cite{logothetidis2014handbook}. Thus, there is a tremendous amount of research being conducted in the field of academia (Oxford Centre for Nonlinear Partial Differential Equations, Lu Research Group the University of Texas at Austin, Flexible Electronics and Display Center Arizona State University) as well as in the commercial sector (LG Electronics \cite{yang1998liquid, song2010electronic}, Samsung Group \cite{jin200965, kim2001portable, kim2012folder}).\\

Elastic foundation models are also used in the study of other mechanical concepts such as the buckling of stiff films bound to compliant substrates under compression \cite{audoly2008buckling1, audoly2008buckling2, audoly2008buckling3} (which are considered to be important in designing of structural sandwich panels \cite{allen2013analysis}) and in the study of crack patterns in thin films subjected to equi-biaxial residual tensile stress \cite{xia2000crack} (which are considered to be important in the study of spiral cracks in thin brittle adhesive layers bonding glass plates together, due to environmental interactions with the residual stress state resulting from processing \cite{dillard1994spiral}). For such applications, Baldelli and Bourdin \cite{Andres} analyse the asymptotic behaviour of bonded thin elastic structures (i.e. films and plates) on elastic foundations. The work is presented as the first attempt at providing a rigorous derivation of these heuristic models from three-dimensional elasticity.\\

In our reading, we show that Baldelli and Bourdin's  work \cite{Andres} is only valid when describing the behaviour of films bonded to elastic pseudo-foundations, where Poisson's ratios of the both bodies are in between $-1$ and $0$ or in between $0$ and $\frac12$ (where both Poisson's ratios are sufficiently away from $0$ and $\frac12$), and an the asymptotic condition that is different to what the authors present. We also show that, for all Poisson's ratios, the authors' phase diagram is four-dimensional, but not two-dimensional as the authors present. Also, due to the Poisson's ratio dependence, the asymptotic scalings that the authors present are insufficient to derive their proposed models. Furthermore, the authors' scaling of the displacement field implies that their method cannot be applicable to films (or strings) with planar loading, unless the normal displacement is zero. Finally, by deriving a Winkler foundation type solution for a plate supported by an elastic pseudo-foundation via the method implied by the authors, we show that the authors' method cannot be applied to plates due to the structure of the overlying body (i.e. the limits of integration of the plate) and the foundation (i.e. planar-stress free condition of the foundation), unless planar displacement field is identically zero. Despite the limitations of the authors' work, we conclude by showing that unlike the classical derivation of the Winkler foundation equations, Baldelli and Bourdin's approach does not violate the volume conversation laws or violate the governing equations of mathematical elasticity.

\subsection{Baldelli and Bourdin's Work}

Baldelli and Bourdin \cite{Andres} perform an asymptotic study to explore the different asymptotic regimes reached in the limit as the thickness of the overlying thin body goes to zero: for varying thickness of the foundation and stiffness ratios. They give a two-dimensional phase diagram to visualise the asymptotically reduced dimension models as a function of two relevant parameters. Two of the major presented results are the identification of the regime of films over in-plane elastic foundations and the identification of the regime of plates over out-of-plane elastic foundations.\\

Baldelli and Bourdin \cite{Andres} begin by classifying the study of thin objects on elastic foundations as Winkler foundations and asserting that the derivation of the Winkler foundations equations must be done with rigorous asymptotic analysis, possibly as described by their publication. However, the authors' assertion is false. Winkler foundation is a very specific mathematical problem, where an elastic body is supported unilaterally on a bed of continuously distributed springs with a foundation modulus $\mathscr K_0$, and where the surface of the foundation is lubricated so that no tangential forces can develop (see section 5.5 of Kikuchi and Oden \cite{Kikuchi} and section 10.4.1 of Ding \emph{et al.} \cite{ding2006elasticity}). Thus, Winkler foundation type problem is a boundary condition that exists regardless of the elastic properties of the elastic body or the bed of springs (see equation 5.111 of Kikuchi and Oden \cite{Kikuchi}). Often in the engineering community, Winkler foundation equations are used to describe the behaviour of beams and plates on elastic foundations with infinite depth \cite{hetenyi1971beams}, with complete disregard to understanding why the Winkler foundation equations are applicable to modelling such problems. Thus, the work of the authors may have intended to be used in justifying the use of Winkler foundation equation in modelling such problems. \\

\begin{figure}[!h]
\centering
\includegraphics[width=0.75\linewidth]{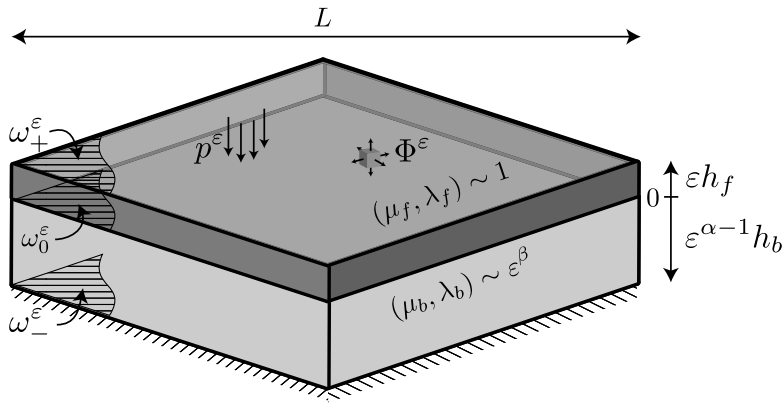}
\caption{'Film on the bonding layer' \label{AndresPic01} \cite{Andres}.}
\end{figure}
%trim=0cm 19.5cm 8cm 0cm, clip = true,

The core idea behind Baldelli and Bourdin \cite{Andres} is as follows. Consider a thin overlying elastic body (which the authors called the film or the membrane) with a constant thickness $\varepsilon h_f$, bonded to an elastic foundation (which the authors called the bonding layer) with a constant thickness $\varepsilon^{\alpha -1} h_b$, where the displacement of the bottom of the bonding layer is zero, i.e. displacement field of the bonding layer satisfies  zero-Dirichlet boundary condition at its lowest boundary (see figure \ref{AndresPic01}). The parameter $\varepsilon$ is considered to be a small constant and $\alpha$ is yet to be determined. In their analysis, the authors assume that there exists a common asymptotic behaviour between the elastic properties (the Young's modulus and the Poisson's ratio) of the overlying body and the bonding layer. To be more precise, the authors assume that both first and second Lam\'{e}'s parameters of the overlying body are of the same order, i.e. $\lambda_f\sim \mu_f$, both first and the second Lam\'{e}'s parameters of the bonding layer are of the same order, i.e. $\lambda_b\sim \mu_b$ (see figure \ref{AndresPic01} or figure 1 of Baldelli and Bourdin \cite{Andres}), both Poisson's ratios of the overlying body and the bonding layer are of the same order, i.e. $\nu_f\sim \nu_b$ (see hypothesis 2 of Baldelli and Bourdin \cite{Andres}), and both Poisson's ratios of the overlying body and the bonding layer are the same sign, i.e. $\nu_b/\nu_f > 0 $ (see remark 2 of Baldelli and Bourdin \cite{Andres}). Now, these conditions result in the following,
\begin{align*}
-1<\nu_f &\approx \nu_b < 0~~\text{or}\\
0<\nu_f &\approx \nu_b < \frac12~,
\end{align*}
given that both Poisson's ratios are sufficiently away from $0$ and $\frac12$. To be more precise, the conditions $\lambda_f\sim \mu_f$ and $\lambda_b\sim \mu_b$ imply that $\lambda_f = c_f\mu_f$ and $\lambda_b= c_b \mu_b$ for some $c_f, c_b \sim 1$ constants, and thus, $\nu_f = \frac12 (1+c_f)^{-1}c_f$ and $\nu_b = \frac12 (1+c_b)^{-1}c_b$. Furthermore, the conditions $\nu_f\sim \nu_b$ and $\nu_b/\nu_f > 0 $ imply the following, 
\begin{align*}
-\frac{|c_f|}{2(1-|c_f|)} &= \nu_f \approx \nu_b = -\frac{|c_b|}{2(1-|c_b|)}~~\text{or}\\
 \frac{|c_f |}{2(1+|c_f|)}&= \nu_f \approx \nu_b =  \frac{|c_b|}{2(1+|c_b|)}~.
\end{align*}
Thus, one gets the condition $-1<\nu_f \approx \nu_b < 0$, sufficiently away from $0$, or the antithetical condition $0<\nu_f \approx \nu_b < \frac12$, sufficiently away from $0$ and $\frac12$. However, Poisson's ratio of an object can vary strictly between $-1$ and $\frac{1}{2}$ \cite{howell2009applied} and different materials have different Poisson's ratios, and thus, Baldelli and Bourdin's assertion \cite{Andres} cannot hold in general. For example, assume that the bonding layer's Poisson's ratio is $\frac{1}{4}$ and the overlying body's Poisson's ratio is infinitesimally small, i.e. $\varepsilon$, and thus, one finds $\nu_b/\nu_f \sim \varepsilon^{-1}$, which violates the authors' assumption.\\

\begin{figure}[!h]
\centering
\includegraphics[width=0.75\linewidth]{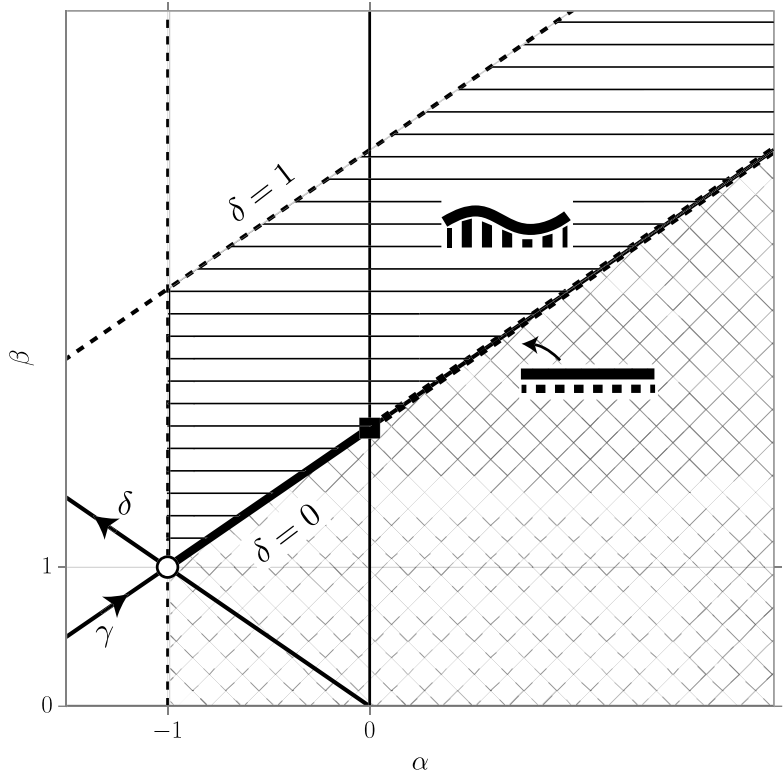}
\caption{The phase plane: `The square-hatched region represents systems behaving as ``rigid'' bodies, under the assumed scaling hypotheses on the loads. Along the open half line (displayed with a thick solid and dashed stroke) $(\delta, 0)$, $\delta> 0$ lay systems whose limit for vanishing thickness leads to a ``membrane over in-plane elastic foundation'' mode ... The solid segment $0 < \gamma <1$ (resp. \emph{dashed open line} $\gamma>1$) is related to systems in which bonding layer is thinner (resp. \emph{thicker}) than the film, for $\gamma = 1$ (\emph{black square}) their thickness is of the same order of magnitude. All systems within the horizontally hatched region $\gamma >0$, $0 < \delta\leq 1$, $\delta > \gamma$ behave, in the vanishing thickness limit, as ``plates over out-of-plane elastic foundation''.' \cite{Andres}. Note that $\gamma = \frac{1}{2}(1+q-\alpha)$, $\delta = \frac{1}{2}(\alpha+q-3)$ and $E_b/E_f \sim \varepsilon^q$. \label{AndresPic02}}
\end{figure}
%trim=0cm 15.5cm 7cm 0cm, clip = true, 

As a result of the restrictive nature of Poisson's ratios of the authors' analysis, they assert that all asymptotic behaviour of the overlying bonded body on an elastic bonding layer can be expressed on a two-dimensional phase diagram (see figure \ref{AndresPic02}). However, this cannot hold in general as the phase diagram is four-dimensional due to the following four asymptotic scalings, $$\left\{\varepsilon^{\alpha -2} \frac{h_b}{h_f},~\frac{\lambda_b}{\mu_f},~\frac{\mu_b}{\mu_f},~\frac{2\nu_f}{(1-2\nu_f)}\right\},$$ for all Poisson's ratios. The only way one may collapse the dimensionality of the phase diagram is by assuming that one is only considering Poisson's ratios with the very specific values $-1<\nu_f \approx \nu_b < 0$ or $0<\nu_f \approx \nu_b < \frac12$, given that both Poisson's ratios are sufficiently away from $0$ and $\frac12$.\\

While describing the rigorous asymptotic analysis, the authors asymptotically rescale the displacement field as $u_\varepsilon = \boldsymbol(\varepsilon u_\varepsilon^1,~\varepsilon u_\varepsilon^2,~ u_\varepsilon^3\boldsymbol)_\text{E}$ (see equation 9 of Baldelli and Bourdin \cite{Andres}), which is implied by hypothesis 1 of the publication. If one defines the displacement field as described, then the only physical interpretation is that the planar displacement field $\boldsymbol(\varepsilon u^1, ~\varepsilon u^1,~0\boldsymbol)$ is infinitesimally small relative to the normal displacement  field $\boldsymbol(0,~0,~u^3\boldsymbol)$, and such scaling results in only plate like solutions. Thus, the authors' claim regarding leading order solution for a film bonded to an elastic foundation is possible is in direct conflict with their choice of the scalings for the displacement field.\\

As an example of their analysis, the authors put forward a model for an overlying film (defined as a membrane) with a very high Young's modulus (i.e. stiff) bonded to an elastic foundation (see theorem 1 of Baldelli and Bourdin \cite{Andres}). With rigorous mathematics, the authors show that there exists a unique solution in $\boldsymbol H^1(\omega)$, where $\omega \subset \mathbb{R}^2$ is the unstrained contact surface between the film and the bonding layer (see section 3.2 of Baldelli and Bourdin \cite{Andres}). Beneath the authors' analysis, the method in which the authors use to derive the governing equations is simple. Below, we describe in detail the method used by the authors to derive the energy functional of a film bonded to an elastic foundation. However, we omit the authors' restrictive scalings of the displacement field (see equation 9 of Baldelli and Bourdin \cite{Andres}) and Poisson's ratios (see figure 1 of Baldelli and Bourdin \cite{Andres}), and the insufficient asymptotic condition $E_fh_f\gg E_bh_b$, where $E_f$ and $E_b$ are the respective Young's moduli of the film and the bonding layer (see definition of $\delta$ of Baldelli and Bourdin \cite{Andres}). Note that \emph{Einstein's summation notation} (see section 1.2 of Kay \cite{kay1988schaum}) is assumed  throughout,  bold symbols signify that we are dealing with vector and tensor fields, we regard the indices $i,j,k,l \in \{1,2,3\}$ and  $\alpha,\beta,\gamma,\delta \in \{1,2\}$, and the coordinates $(x^1,x^2,x^3)=(x,y,z)$, unless it is strictly states otherwise.\\

Consider an overlying film with a Poisson ratio $\nu_f$, a Young's modulus $E_f$ and a thickness $h_f$, and a bonding layer (i.e. a foundation) with a Poisson ratio $\nu_b$,  a Young's modulus $E_b$ and a thickness $h_b$. Now, define the displacement field of the film by the following, $$\boldsymbol u_f(\boldsymbol u) = \boldsymbol (u^1(x^1,x^2), ~u^2(x^1,x^2), ~0 \boldsymbol)$$ and the displacement field of the foundation by the following,
\begin{align}
\boldsymbol u_b(\boldsymbol u) =\left[1+\frac{1}{h_b}x^3\right]\boldsymbol (u^1(x^1,x^2), ~u^2(x^1,x^2), ~0\boldsymbol)~, \label{filmu}
\end{align}
where $x^\alpha \in \omega$ and $x^3 \in (-h_b, 0)$. One can see that the displacement field of the foundation satisfies zero-Dirichlet boundary condition (i.e. $\boldsymbol u_b(\boldsymbol u) |_{x^3=- h_b}= \boldsymbol 0$), and placement field of the film and the foundation are continuous at the contact region (i.e. $\boldsymbol u_b(\boldsymbol u)|_{x^3=0} = \boldsymbol u_f(\boldsymbol u) |_{x^3=0}$). Now,  the energy functional of the system can be expressed as follows,
\begin{align*}
J(\boldsymbol u) = ~& \int_0^{h_f}\int_\omega \left[\frac12 A^{\alpha\beta\gamma\sigma}_f \epsilon_{\alpha\beta}(\boldsymbol u_f(\boldsymbol u)) \epsilon_{\gamma\delta}(\boldsymbol u_f(\boldsymbol u)) -f^\alpha u_\alpha\right]d\omega dx^3 \\
& + \int^0_{-h_b}\int_\omega \frac12 A^{ijkl}_b \epsilon_{ij}(\boldsymbol u_b(\boldsymbol u) ) \epsilon_{kl}(\boldsymbol u_b(\boldsymbol u)) ~d\omega dx^3~,
\end{align*}
where $\epsilon_{ij}(\boldsymbol v) = \frac{1}{2}(\partial_i v_j + \partial_j v_i)$ is the linearised Green-St Venant strain tensor, $\boldsymbol (f^1,f^2,0\boldsymbol)$ is an external planar force density field,
\begin{align*}
A^{\alpha\beta\gamma\delta}_f= \frac{2\mu_f\lambda_f }{(\lambda_f+2\mu_f)}\delta^{\alpha\beta}\delta^{\gamma\delta} + \mu_f\left( \delta^{\alpha\gamma}\delta^{\beta\delta}+\delta^{\alpha\delta}\delta^{\beta\gamma}\right)
\end{align*}
is the elasticity tensor of the film (or the plate in subsequent analysis), 
\begin{align*}
A^{ijkl}_b= \lambda_b \delta^{ij}\delta^{kl} + \mu_b\left(\delta^{ik}\delta^{jl}+\delta^{il}\delta^{jk}\right)
\end{align*}
is the elasticity tensor of the foundation, and were $\delta^j_i$ is the Kronecker delta. Note that 
\begin{align*}
\lambda_f &=\frac{\nu_fE_f}{(1+\nu_f)(1 -2\nu_f)}~~\text{and}\\
\mu_f &= \frac{E_f}{2(1+\nu_f)}
\end{align*}
are the first and the second Lam\'{e}'s parameters of the film respectively, and
\begin{align*}
\lambda_b &=\frac{\nu_bE_b}{(1+\nu_b)(1 -2\nu_b)}~~\text{and}\\
\mu_b &= \frac{E_b}{2(1+\nu_b)}
\end{align*}
are the first and the second Lam\'{e}'s parameters of the foundation respectively. Due to Poisson's ratio dependence, one comes to the conclusion that 
\begin{align}
\left\{\Lambda_f h_f \sim \mu_b\frac{\mathrm{meas}(\omega;\mathbb{R}^2)}{h_b},~ \Lambda_f h_f \gg (\lambda_b+2\mu_b) h_b\right\} \label{filmscale}
\end{align}
is the only possible asymptotic scaling that allows any valid leading-order governing equations (i.e. problems that allow traction), where 
\begin{align*}
\Lambda_f &=4\mu_f\frac{(\lambda_f+\mu_f)}{(\lambda_f+2\mu_f)}\\
& = \frac{E_f}{(1+\nu_f)(1 -\nu_f)}~,
\end{align*}
and $\mathrm{meas}(\cdot;\mathbb{R}^n)$ is the standard Lebesgue measure in $\mathbb{R}^n$. To be more precise, scaling (\ref{filmscale}) is the asymptotic scaling that allows the following conditions,
\begin{align}
\int_0^{h_f}\int_\omega \frac12 A^{\alpha\beta\gamma\delta}_f \epsilon_{\alpha\beta}(\boldsymbol u_f(\boldsymbol u)) \epsilon_{\gamma\delta}(\boldsymbol u_f(\boldsymbol u)) ~d\omega dx^3 \approx & \nonumber\\
\int^0_{-h_b}\int_\omega A^{\alpha3\beta3}_b \epsilon_{\alpha3}(\boldsymbol u_b(\boldsymbol u)) \epsilon_{\beta 3}(\boldsymbol u_b(\boldsymbol u)) &~d\omega dx^3~~ \text{and} \nonumber \\
\int_0^{h_f}\int_\omega \frac12 A^{\alpha\beta\gamma\delta}_f \epsilon_{\alpha\beta}(\boldsymbol u_f(\boldsymbol u)) \epsilon_{\gamma\delta}(\boldsymbol u_f(\boldsymbol u)) ~d\omega dx^3  \gg & \nonumber \\
\int^0_{-h_b}\int_\omega \frac12 A^{\alpha\beta\gamma\sigma}_b \epsilon_{\alpha\beta}(\boldsymbol u_b(\boldsymbol u)) \epsilon_{\gamma\delta}(\boldsymbol u_b(\boldsymbol u)) &~d\omega dx^3~.\label{asymptoticA}
\end{align}
To see why relation (\ref{asymptoticA}) implies the condition $\Lambda_f h_f \gg (\lambda_b+2\mu_b) h_b$, please consult the proof of theorems 3.9-1 and theorem 4.4-1of Ciarlet \cite{ciarlet2005introduction}.\\

Now, with a little more asymptotic analysis, one can express the leading-order terms of the energy functional of a film bonded to an elastic pseudo-foundation as follows,
\begin{align}
J(\boldsymbol u) = \frac12 h_f\int_\omega\bigg[ \frac{2\mu_f\lambda_f}{(\lambda_f+2\mu_f)} \epsilon_\alpha^\alpha(\boldsymbol u) \epsilon_\beta^\beta(\boldsymbol u) &+2\mu_f\epsilon_\alpha^\beta(\boldsymbol u) \epsilon_\beta^\alpha(\boldsymbol u) \nonumber \\
&+ \left(\frac{\mu_b}{h_f h_b}\right) u_\alpha u^\alpha - 2f^\alpha u_\alpha \bigg] d\omega~. \label{andresRight01}
\end{align}
If $\omega \subset \mathbb{R}^2$ is a connected bounded plane with a Lipchitz-continuous boundary $\partial\omega$, and $f^\alpha \in L^1(\omega)$, then there exist a unique minimiser $\boldsymbol(u^1, u^2\boldsymbol ) \in \boldsymbol H^1(\omega)$ to equation (\ref{andresRight01}) (see section 1.5 of Ciarlet \cite{Ciarlet1997theory}, in particular Korn's inequality on a surface without boundary conditions, but with $u^3=0$). In particular, this unique minimiser is also a critical point in $\boldsymbol H^1(\omega)$ (see section 1.5 of Badiale and Serra \cite{badiale2010semilinear}). Note that $L^k(\cdot)$ are the standard $L^k$-Lebesgue spaces and $H^k(\cdot)$ are the standard $W^{k,2}(\cdot)$-Sobolev spaces (see section 5.2.1 of Evans \cite{Evans}). Also note that we called the elastic foundation as an elastic pseudo-foundation as now the displacement field of the foundation (\ref{filmu}) is grossly oversimplified.\\

Note that the asymptotic condition $\Lambda_f h_f \sim h_b^{-1}\mu_b\mathrm{meas}(\omega;\mathbb{R}^2)$ implies that arbitrarily increasing the Young's modulus of the film or arbitrary decreasing the thickness of the film, relative to the foundation (implied by the authors' asymptotic scaling $E_f h_f \gg E_b h_b$), may not generate the most accurate solution. This is demonstrated numerically by Jayawardana \cite{jayawardana2016mathematical} (see section 3.6 and 3.7 of Jayawardana \cite{jayawardana2016mathematical}) in curvilinear coordinates (i.e. a shell bonded to an elastic foundation), as the author observe optimal values for the Young's moduli, the Poisson's ratios and thickness of each elastic body (and the curvature of the contact region), where the error of planar solution (with respect to the solution implied by standard linear elasticity) attains a minimum.\\

Baldelli and Bourdin \cite{Andres} go further with their approach to derive a set of governing equations to describe the behaviour of a stiff plate bonded to an elastic foundation (see theorem 2 of Baldelli and Bourdin \cite{Andres}). For this, the authors define the energy functional as follows,
\begin{align}
J(\boldsymbol u) = & \int_0^{h_f}\int_\omega \left[\frac12 A^{\alpha\beta\gamma\sigma}_f \epsilon_{\alpha\beta}(\boldsymbol u_f(\boldsymbol u)) \epsilon_{\gamma\delta}(\boldsymbol u_f(\boldsymbol u)) -f^i u_i\right]d\omega dx^3 \nonumber \\
&+ \int^0_{-h_b}\int_\omega \frac{4\mu_b(\lambda_b+\mu_b)}{\lambda_b+2\mu_b} \epsilon^3_3(\boldsymbol u_b(\boldsymbol u)) \epsilon^3_3(\boldsymbol u_b(\boldsymbol u)) ~d\omega dx^3~, \label{andresWrongXXX}
\end{align}
where in the theorem, respective displacement fields of the plate and the foundation are defined  as follows,
\begin{align*}
\boldsymbol u_f(\boldsymbol u)  = ~&\boldsymbol(u^1(x^1,x^2), ~u^2(x^1,x^2),~ u^3(x^1,x^2)\boldsymbol)~~\text{and}\\
\boldsymbol u_b(\boldsymbol u) = ~&\boldsymbol(u^1(x^1,x^2),~ u^2(x^1,x^2),~ u^3(x^1,x^2)\boldsymbol) \nonumber\\
&+ \left[x^3+ h_b\right]\boldsymbol(\partial^1 u_3(x^1,x^2),~\partial^2 u_3(x^1,x^2),~ 0\boldsymbol)~,
\end{align*}
and in the proof of the theorem, respective displacement fields of the plate and the foundation are defined as follows,
\begin{align*}
\boldsymbol u_f(\boldsymbol u)  =~&\boldsymbol(u^1(x^1,x^2),~ u^2(x^1,x^2),~ u^3(x^1,x^2)\boldsymbol) \nonumber\\
& - \left[x^3-\frac12 h_f\right]\boldsymbol(\partial^1 u_3(x^1,x^2),~ \partial^2 u_3(x^1,x^2),~ 0\boldsymbol)~~\text{and}\\
\boldsymbol u_b(\boldsymbol u)  = ~&\boldsymbol(u^1(x^1,x^2), ~u^2(x^1,x^2) ,~ [h_b +x^3] u^3(x^1,x^2)\boldsymbol)~.
\end{align*}

Unfortunately, the authors' theorem 2 is erroneous (see equation 18-22 of Baldelli and Bourdin \cite{Andres}) and their proof theorem 2 (see section 3.3 of Baldelli and Bourdin \cite{Andres}) is inapplicable to the authors' theorem. For example, one can clearly see that there are two conflicting definitions for the displacement fields, both definitions of the displacement fields are not continuous at the contact region, i.e. $\boldsymbol u_b(\boldsymbol u)|_{x^3=0} \neq \boldsymbol u_f(\boldsymbol u) |_{x^3=0}$, and both definition of the displacement field of the foundation does not satisfy the zero-Dirichlet boundary condition, i.e. $\boldsymbol u_b(\boldsymbol u) |_{x^3=- h_b}\neq \boldsymbol 0$. Also, the authors' asymptotic scalings from their proof of theorem 2 does not result in equation (\ref{andresWrongXXX}). To be precise, consider a plate bonded to an elastic foundation with a constant thickness, such that the lower surface of the foundation satisfies the zero-Dirichlet boundary condition, where the displacement fields of the plate and the foundation can be express respectively as follows,
\begin{align*}
\boldsymbol u_f(\boldsymbol u)  =~&\boldsymbol(u^1(x^1,x^2),~ u^2(x^1,x^2),~ u^3(x^1,x^2)\boldsymbol) \nonumber\\
& - x^3\boldsymbol(\partial^1 u_3(x^1,x^2),~ \partial^2 u_3(x^1,x^2),~ 0\boldsymbol)~~\text{and}\\
\boldsymbol u_b(\boldsymbol u)  = ~&\left[1+\frac{1}{h_b}x^3\right] \boldsymbol(u^1(x^1,x^2), ~u^2(x^1,x^2) , ~u^3(x^1,x^2)\boldsymbol)~.
\end{align*}
Thus, the energy functional of this two elastic bodies can be expressed as follows,
\begin{align}
J(\boldsymbol u) = & \int_0^{h_f}\int_\omega \left[\frac12 A^{\alpha\beta\gamma\sigma}_f \epsilon_{\alpha\beta}(\boldsymbol u_f(\boldsymbol u)) \epsilon_{\gamma\delta}(\boldsymbol u_f(\boldsymbol u)) -f^i u_i\right]d\omega dx^3 \nonumber \\
&+ \int^0_{-h_b}\int_\omega \frac12 A^{ijkl}_b \epsilon_{ij}(\boldsymbol u_b(\boldsymbol u)) \epsilon_{kl}(\boldsymbol u_b(\boldsymbol u)) ~d\omega dx^3~. \label{andresWrongXXX2}
\end{align}
However, there exists no asymptotic scaling (what Baldelli and Bourdin present \cite{Andres} or otherwise) such that the leading order terms of equation (\ref{andresWrongXXX2}) would result in equation (\ref{andresWrongXXX}), unless the planar displacement field is an order of magnitude smaller than the normal displacement, i.e. $u^3 \gg u^\alpha$, $\forall ~\alpha\in\{1,2\}$. To see this more clearly, integrate equation  (\ref{andresWrongXXX2}) explicitly in $x^3$ dimension to find the following,
\begin{align}
J(\boldsymbol u) =  J_\text{plate}(\boldsymbol u) + J_\text{foundation}(\boldsymbol u)~, \label{andresWrongXXX3}
\end{align}
where 
\begin{align}
J_\text{plate}(\boldsymbol u) =  \int_\omega \bigg[\frac12 A^{\alpha\beta\gamma\sigma}_f  \bigg(&
 h_f \epsilon_{\alpha\beta}(\boldsymbol u) \epsilon_{\gamma\delta}(\boldsymbol u)\nonumber\\
& - \frac12 h_f^2\big(\partial_{\alpha\beta}u^3\epsilon_{\gamma\delta}(\boldsymbol u) + 
\epsilon_{\alpha\beta}(\boldsymbol u) \partial_{\gamma\delta}u^3\big) \nonumber\\
& + \frac13 h_f^3\partial_{\alpha\beta}u^3\partial_{\gamma\delta}u^3\bigg) -f^i u_i \bigg]d\omega \label{andresWrongXXX4}
\end{align}
and
\begin{align*}
J_\text{foundation}(\boldsymbol u) = \int_\omega \frac12 \bigg[ & \lambda_b \left(\frac13 h_b \epsilon^\alpha_\alpha (\boldsymbol u) \epsilon^\beta_\beta(\boldsymbol u) + u^3\epsilon^\alpha_\alpha (\boldsymbol u) + \frac{1}{h_b}u^3u_3 \right) \\
&+ 2\mu_b\bigg(\frac13 h_b \epsilon_{\alpha\beta}(\boldsymbol u) \epsilon_{\gamma\delta}(\boldsymbol u) \\
&+ \frac12 \big(\frac{1}{h_b} u^\alpha u_\alpha + u^\alpha\partial_\alpha u^3 + \frac13h_b\partial_\alpha u^3\partial^\alpha u_3\big) \nonumber\\
&+ \frac{1}{h_b} u^3u_3\bigg) \bigg]~d\omega~.
\end{align*}
As the reader can see that there exist no asymptotic scaling that one can apply to equation (\ref{andresWrongXXX3}) to get a leading order equation of the form (\ref{andresWrongXXX}), unless $u^\alpha =0$, $\forall ~\alpha\in\{1,2\}$. Note that in standard linear plate theory, if the mid-plane of the plate is located at $x^3=0$, then the limits of integration for a plate with thickness $h_f$ are $x^3\in[-\frac12 h_f, \frac12 h_f]$, which results in only the terms $\epsilon_{\alpha\beta}(\boldsymbol u) \epsilon_{\gamma\delta}(\boldsymbol u)$ and $\partial_{\alpha\beta}u^3\partial_{\gamma\delta}u^3$ in the plate's energy functional. However, as a result of the authors' limits of integration, i.e. $x^3\in[0, h_f]$, we see the product $\epsilon_{\alpha\beta}(\boldsymbol u) \partial_{\gamma\delta}u^3$ appearing in equation (\ref{andresWrongXXX4}).\\

We now derive a solution for a plate supported by an elastic foundation, in accordance with the techniques implied by the proof of theorem 2, but with mathematical rigour. Just as before, we omit the authors' restrictive scalings of the displacement field (see equation 9 of Baldelli and Bourdin \cite{Andres}) and Poisson's ratios (see figure 1 of Baldelli and Bourdin \cite{Andres}), and the insufficient asymptotic condition $E_fh_f\gg E_bh_b$ (see definition of $\delta$ of Baldelli and Bourdin \cite{Andres}). Also, we only include leading order terms for the sake of readability.\\

Assume that we are considering Winkler foundation type problem, and let $T^i_j(\boldsymbol v) = A_b^{ijkl}\epsilon_{kl}(\boldsymbol v)$ be the second Piola-Kirchhoff stress tensor and let $\boldsymbol v \in \omega\times[-h_b, 0)$ be the displacement field of the foundation. By definition, Winkler foundations cannot admit planar stress (see section 5.5 of Kikuchi and Oden \cite{Kikuchi}), and thus, implies  $T^\alpha_\beta(\boldsymbol v) = 0$, $\forall ~\alpha, \beta \in\{1,2\}$. Now, use these conditions to modify the elasticity tensor and the displacement field of the foundation, i.e. use the conditions $\epsilon^\alpha_\beta(\boldsymbol v) = 0$, $\forall ~\alpha, \beta \in\{1,2\}$ and $\epsilon^\alpha_\alpha(\boldsymbol v) = -(\lambda_b+\mu_b)^{-1} \lambda_b \epsilon^3_3(\boldsymbol v)$ to obtain  the following,
\begin{align*}
T^j_3(\boldsymbol v(\boldsymbol u_b)) = \frac{\lambda_b\mu_b}{(\lambda_b+\mu_b)}\epsilon^3_3(\boldsymbol v(\boldsymbol u_b)) \delta^j_3+ 2\mu_b\epsilon^j_3(\boldsymbol v(\boldsymbol u_b))~,
\end{align*}
and thus, one gets $T^{j3}(\boldsymbol u_b) = A^{j3k3}_{b^\text{new}}\epsilon_{k3}(\boldsymbol u_b)$, where
\begin{align*}
A^{jikl}_{b^\text{new}} = \frac{\lambda_b\mu_b}{(\lambda_b+\mu_b) }\delta^{ji}\delta^{kl} + \mu_b(\delta^{jk}\delta^{il}+\delta^{jl}\delta^{ik})
\end{align*}
is the new elasticity tensor of the foundation and $\boldsymbol u_b$ is the new the displacement field of the foundation that satisfies the conditions $\epsilon^\alpha_\beta(\boldsymbol u_b) = 0$, $\forall ~\alpha, \beta \in\{1,2\}$. Now, seek a displacement field $\boldsymbol u_b$ of the form such that $\epsilon^\alpha_\beta(\boldsymbol u_b) = 0$, $\epsilon^j_3(\boldsymbol u_b) \neq 0$ and $u_b^3(x^1,x^2, -h_b)=0$, and thus, one finds that the displacement field of the foundation can be expressed as follows,
\begin{align*}
\boldsymbol u_b (\boldsymbol u) =\left[1+\frac{1}{h_b}x^3\right] \boldsymbol (0,~0,~u^3(x^1,x^2)\boldsymbol)~,
\end{align*}
where $x^3 \in (0,-h_b)$. Now, assume that there is an overlying plate supported by the foundation. As the displacement field at the contact region must be continuous, one finds that the displacement field of the plate can be expressed as follows,
\begin{align*}
\boldsymbol u_f (\boldsymbol u) = \boldsymbol( -x^3\partial^1 u_3(x^1,x^2),~ -x^3\partial^2 u_3(x^1,x^2), ~u^3(x^1,x^2)\boldsymbol)
\end{align*}
where $x^3 \in [0,h_f)$. Due to Poisson's ratio dependence, one comes to the conclusion that 
\begin{align}
\left\{\Lambda_f h_f^3\sim \mathscr K_b (\mathrm{meas}(\omega;\mathbb R^2))^2, ~ \Lambda_f h^3_f \gg h_b \mu_b\mathrm{meas}(\omega;\mathbb R^2)\right\} \label{platescale}
\end{align}
is the only possible asymptotic scaling that allows any valid governing equations (i.e. that allows Winkler foundation type problems), where 
\begin{align*}
\mathscr K_b &=  \frac{\mu_b(3\lambda_b+2\mu_b)}{h_b(\lambda_b+\mu_b)}\\
& = \frac{E_b}{h_b}
\end{align*}
is the foundation modulus. To be more precise, scaling (\ref{platescale}) is the asymptotic scaling that allows the following conditions,
\begin{align}
\int_0^{h_f}\int_\omega \frac12 A^{\alpha\beta\gamma\delta} _f\epsilon_{\alpha\beta}(\boldsymbol u_f(\boldsymbol u)) \epsilon_{\gamma\delta}(\boldsymbol u_f(\boldsymbol u) ) ~d\omega dx^3 \approx & \nonumber\\
 \int^0_{-h_b}\int_\omega \frac12 A^{3333}_{b^\text{new}} \epsilon_{33}(\boldsymbol u_b(\boldsymbol u)) \epsilon_{33}(\boldsymbol u_b(\boldsymbol u)) &~d\omega dx^3~~ \text{and}  \label{asymptoticB}\\
\int_0^{h_f}\int_\omega \frac12 A^{\alpha\beta\gamma\delta}_f \epsilon_{\alpha\beta}(\boldsymbol u_f(\boldsymbol u)) \epsilon_{\gamma\delta}(\boldsymbol u_f(\boldsymbol u) ) ~d\omega dx^3  \gg &\nonumber \\
 \int^0_{-h_b}\int_\omega A^{\alpha3\beta3}_{b^\text{new}}\epsilon_{\alpha3}(\boldsymbol u_b(\boldsymbol u)) \epsilon_{\beta 3}(\boldsymbol u_b(\boldsymbol u)) &~d\omega dx^3~.\nonumber
\end{align}
To see why relation (\ref{asymptoticB}) implies the condition $\Lambda_f h_f^3\sim \mathscr K_b (\mathrm{meas}(\omega;\mathbb R^2))^2$, please consult the proof of theorems 3.9-1 and theorem 4.4-1of Ciarlet \cite{ciarlet2005introduction}.\\

Now, with a little more asymptotic analysis, one can express the leading-order terms of the energy functional of a plate supported by an elastic pseudo-foundation as follows,
\begin{align}
J(\boldsymbol u) = \frac{1}{2}h_f \int_\omega\left[ \frac13 \Lambda_f h_f^2 \Delta u_3 \Delta u^3 + \left(\frac{\mathscr K_b}{h_f}\right) u_3 u^3 - 2f^3u_3\right]d\omega~, \label{AndresRight}
\end{align}
where $\Delta$ is the scalar Laplacian in the Euclidean plane and $f^3$ is an external normal force density. As the reader can clearly see that equation (\ref{andresWrongXXX}) is different to equation (\ref{AndresRight}), i.e. there exist clear discrepancies between what Baldelli and Bourdin's  theorem 2 \cite{Andres} and the method implied by the proof of theorem 2. Note that we used the condition $\partial_\beta u^3|_{\partial\omega} =0$, $\forall~\beta \in\{1,2\}$, to obtain equation (\ref{AndresRight}), where $\boldsymbol n$ is the unit outward normal to the boundary of the plate, $\partial\omega$. \\

Equation (\ref{AndresRight}) is a Winkler foundation type problem for a plate that is supported  by a continuous bed of springs with a foundation modulus of $\mathscr K_b$. Furthermore, if $\omega \subset \mathbb{R}$ is an connected  bounded plane with a Lipchitz-continuous boundary $\partial\omega$, $f^3 \in L^1(\omega)$, and $\partial_\beta u^3|_{\partial\omega} =0$, $\forall~\beta \in\{1,2\}$, in a trace sense, then there exists a unique minimiser $u^3\in H^2(\omega)$ to equation (\ref{AndresRight}) (see section 1.5 of Ciarlet \cite{Ciarlet1997theory}, in particular Korn's inequality on a surface without boundary conditions, but with $\boldsymbol(u^1, u^2\boldsymbol ) =\boldsymbol 0$). In particular, this unique minimiser is also a critical point in $ H^2(\omega)$ (see section 1.5 of Badiale and Serra \cite{badiale2010semilinear}).\\

Should one try to derive the Winker foundation equation to this problem via the classical approach, Dillard \emph{et al.}' work  \cite{dillard2018review} (see equation (3) of Dillard \emph{et al.} \cite{dillard2018review}) implies an energy functional of the following form,
\begin{align}
J_{\text{classical}}(\boldsymbol u) = \frac{1}{2}h_f \int_\omega\left[ \frac{1}{12} \Lambda_f h_f^2 \Delta u_3 \Delta u^3 + \left(\frac{\mathscr K_b}{h_f}\right) u_3 u^3 - 2f^3u_3\right]d\omega~, \label{dillard}
\end{align}
where the discrepancy with respect to Baldelli and Bourdin's approach \cite{Andres} (i.e. the $\frac12$ term in equation (\ref{AndresRight}) and the $\frac{1}{12}$ term equation (\ref{dillard}))  is due the limits of integration of the plate, i.e. $[-\frac12 h_f, \frac12 h_f]$ for Dillard \emph{et al.} \cite{dillard2018review} (classical derivation) and $[0, h_f]$ for Baldelli and Bourdin \cite{Andres}. Now, given that the plate resides in the set $\omega\times[-\frac12 h_f, \frac12 h_f]$, if the foundation resides in the set $\omega\times[-h_b, 0)$, then $[-\frac12 h_f, 0)$ is a region of overlap (i.e. it violates the volume conservation laws), and if the foundation resides in the set $\omega\times[-h_b -\frac12 h_f, -\frac12 h_f)$, then then $[-\frac12 h_f, 0)$ is a region of discontinuity (i.e. it violates the governing equations of mathematical elasticity). However, in Baldelli and Bourdin's approach \cite{Andres}, the plate resides in the set $\omega\times[0, h_f]$ and the foundation resides in the set $\omega\times[-h_b, 0)$, and thus, no such such regions of overlap or discontinuities. Thus, unlike the classical approach,  Baldelli and Bourdin's approach \cite{Andres} is consistent with the volume conservation laws the governing equations of mathematical elasticity.\\

Note that, if one is assuming a plate bonded to an elastic foundation (i.e. not a Winkler foundation, and thus, we assume an elasticity tensor of the form $A^{ijkl}_b$ in the foundation), and if one is considering normal displacement to be dominant, then one finds an energy functional of the following form,
\begin{align*}
J(\boldsymbol u) = \frac{1}{2}h_f \int_\omega\left[ \frac13 \Lambda_f h_f^2 \Delta u_3 \Delta u^3 + \left(\frac{\lambda_b+2\mu_b}{h_fh_b}\right) u_3 u^3 - 2f^3u_3\right]d\omega~,
\end{align*}
where the elastic and geometric properties must satisfy the following scaling
\begin{align*}
\left\{\Lambda_f h_f^3\sim (\lambda_b+2\mu_b)\frac{ (\mathrm{meas}(\omega;\mathbb R^2))^2}{h_b}, ~ \Lambda_f h^3_f \gg h_b \mu_b\mathrm{meas}(\omega;\mathbb R^2)\right\},
\end{align*}
which is, again, different to what the authors present.\\

\begin{figure}[!h]
\centering
\includegraphics[width=1\linewidth]{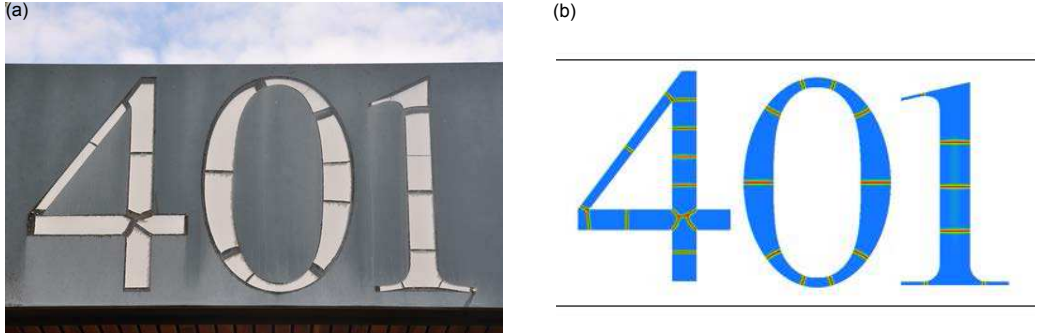}
\caption{(a) `Cracked lettering at Ecole Polytechnique, Palaiseau, France. A vinyl sticker is bonded to an aluminium substrate and exposed to the sun which causes tensile stresses and subsequent cracking.' (b) 'Numerical experiment: nucleation at weak singularities, multiple cracking in the smooth domain, periodic fissuration of slender segments' \cite{baldelli2013fracture}. \label{AndresPic03}}
\end{figure}

Despite above highlighted flaws, the strength of Baldelli and Bourdin's \cite{Andres} work appears lies in the study of overlying bonded films on elastic foundations, which is the subject of study in  the PhD thesis of A. A. L. Baldelli \cite{baldelli2013fracture}\footnote{\href{http://citeseerx.ist.psu.edu/viewdoc/download?doi=10.1.1.402.3946\&rep=rep1\&type=pdf}{ http://citeseerx.ist.psu.edu/viewdoc/download?doi=10.1.1.402.3946\&rep=rep1\&type=pdf}}, where the author use the bonded film model to examine the crack patterns that occurs in thin structures (see page 147 of Baldelli \cite{baldelli2013fracture}). Baldelli \cite{baldelli2013fracture} numerically shows  that, without any `priori' assumptions on the crack geometry, one can capture complex evolving crack patterns in different asymptotic regimes: parallel, sequential, periodic cracking and possible debonding in a uni-axial traction test as well as the appearance of polygonal crack patterns in a two-dimensional equi-biaxial load, and cracking in a geometrically complex domain. One of the perfect examples of the author's work is a comparison against a real life crack pattern and the author's numerical model, where the reader can see from figure \ref{AndresPic03} that the author's numerical result in figure \ref{AndresPic03}-(b) is almost identical to the real crack pattern observed in figure \ref{AndresPic03}-(a). 

\section{Conclusions}

In conclusion, Baldelli and Bourdin's \cite{Andres} work is only valid when describing the behaviour of overlying bonded films on elastic pseudo-foundations (note that the authors' foundation is not an actual elastic foundation as the displacement field of the foundation is grossly over simplified), where Poisson's ratios of the both bodies are in between $-1$ and $0$ or in between $0$ and $\frac12$ (where both Poisson's ratios are sufficiently away from $0$ and $\frac12$), and with the asymptotic condition $\{E_f h_f \sim h_b^{-1}E_b\mathrm{meas}(\omega;\mathbb R^2),~E_f h_f \gg E_b h_b\}$ (where the authors assume only the condition $E_f h_f \gg E_b h_b$ is significant). The authors assert that their asymptotic approach is valid for all elastic properties. However, we mathematically proved that the authors' asymptotic approach is only valid if both Poisson's ratios are in between $-1$ and $0$ or in between $0$ and $\frac12$ (where both Poisson's ratios are sufficiently away from $0$ and $\frac12$). For all Poisson's ratios, the authors' phase diagram is four-dimensional, but not two-dimensional as the authors present. Also, due to the Poisson's ratio dependence, the only scalings that can yield any valid asymptotic solutions are $\{\Lambda_f h_f \sim h_b^{-1}\mu_b\mathrm{meas}(\omega;\mathbb R^2), ~\Lambda_f h_f \gg (\lambda_b+2\mu_b) h_b\}$ for a film that is bonded to an elastic foundation, and $\{\Lambda_f h_f^3\sim \mathscr K_b (\mathrm{meas}(\omega;\mathbb R^2))^2,~\Lambda_f h^3_f \gg h_b \mu_b\mathrm{meas}(\omega;\mathbb R^2)\}$ for a plate that is supported by an elastic foundation, but not $E_fh_f\gg E_bh_b$ as the authors present. The authors' scaling of the displacement field implies that the method cannot be applicable to films (or strings) with planar loading, unless $u^3$ is zero. Finally, by deriving a Winkler foundation type solution for a plate supported by an elastic pseudo-foundation via the method implied by the authors' proof of theorem 2, we showed that the authors' method cannot be applied to plates due to the structure of the overlying body (i.e. limits of integration of the plate) and the foundation (i.e. the planar-stress free condition of the foundation), unless field $\boldsymbol(u^1,u^2\boldsymbol)$ is identically zero.\\

We conclude by noting that the benefit of Baldelli and Bourdin's work \cite{Andres} is that unlike the classical derivation of the Winkler foundation equations, the authors' approach is consistent with the volume conservation laws the governing equations of mathematical elasticity.

\bibliographystyle{./model1-num-names}
\bibliography{CriticaStudyOfBaldellisWork}%
\biboptions{sort&compress}

\end{document}